\documentstyle[12pt]{article}

\begin{document}

\topmargin 0pt
\oddsidemargin 5mm

\setcounter{page}{1}
%\begin{titlepage}
\vspace{2cm}
\begin{center}

{\bf Femtosecond Transversal Deflection of Electron Beams with the Help of 
Laser Beams and Its Possible Applications}\\

\vspace{3mm}
{\large K.A. Ispirian $ \footnote{e-mail: karo@lx2.yerphi.am} $, 
M.K. Ispiryan$^* $  }\\
{\em Yerevan Physics Institute, 375036, Yerevan, Armenia;}\\ 
{\em $^* $ University of Houston, Houston, TX, USA }
\end{center}

\vspace{2mm}

\centerline{\bf{Abstract}}
It is shown that the interaction of an electron beam  with polarized 
electromagnetic wave of laser photons propagating in the same direction 
in a short interaction region results in significant transversal deflection of 
the electrons which can be used for production of femtosecond electron and 
synchrotron radiation beams, for chopping the electron beams and construction 
of laser oscilloscopes measuring femtosecond processes. 


\vspace{10mm}

\indent 
As it is well known (see [1] and references therein) it is difficult to 
accelerate the charged particles by the periodically varying electromagnetic 
field of laser photon beams, especially, in vacuum, because the fields 
are perpendicular to the direction of propagation of photons. Despite the 
fact that there are very strong fields of laser beams with intensity up to
$W \simeq 10^{18} W/cm^2 $ and electric field up to $E \simeq 2.10^{10} V/cm $ 
and optimistic acceleration rates theoretically predicted for various advanced 
methods of particle acceleration, the achieved record acceleration rates are 
less than $\sim 10^{10} $eV/cm at very short distances, and the progress in 
this field is very slow. 

\indent
On the other hand there is a growing interest to the production and study of 
particle and photon bunches with ultrashort time duration connected with 
the advance of new acceleration 
methods, microbunching of particle beams, various radiation mechanism, etc. The 
latest achievements [2-4] of obtaining femtosecond pulses of synchrotron 
radiation with spectral distribution  from infrared to hard X-ray regions 
using femtosecond laser pulses [5] open a possibility to study atomic, 
solid state, chemical and biological processes in a new fundamental time 
scale of $\sim 100$ fs of the order of vibrational period of molecules. 
At present several methods has been developed for the measurement of 
subpicosecond times. The oldest one [6] based on the transverse deflection of 
electron 
beams by the field in RF cavity has been modified in many works and together 
with comercial streak cameras allow to study the bunch length and longitudinal 
particle distribution at time intervals up to $\sim 100 $ fs. However 
these methods are expensive and the theoretical limits of these methods is 
less than 10 fs [6].

\indent
The existing other methods of the study of short processes use the 
properties of various types of coherent radiation the intensity 
of which is proportional to the square of the number 
of the particles in bunches [7,8] when the wavelength of the radiation becomes 
larger than the length of the bunch. With the help of autocorrelation 
information and possible phase data this method can be improved to measure 
time durations down to tens of fs. Moreover as 
it has been shown in [9] the study of the coherent x-ray transition 
radiation of microbunched beams theoretically allows to study processes with  
varying in time intervals down to atosecond, $10^{-18} $s, nevertheless, 
the time 
measurement accuracy achieved by coherent radiation methods is of the order of 
few hundreds of fs. Thus there is a 
need for a method for the measurement of femtosecond times.

\indent 
In this short note it is shown that despite to the difficulties of particle 
acceleration one can successfuly use the laser beams for the transversal 
deflection of electron beams.It is shown that by replacing the deflecting 
RF electromagnetic fields inside the cavities by intense laser fields one 
can construct subfemtosecond oscilloscopes. 

\indent
Let us assume that an electron beam with relativistic factor 
$\gamma = (1 - \beta^2)^{-1/2} = \varepsilon /mc^2 \gg 1 $, 
($m $, $v $, $\varepsilon$ are the mass, velocity and energy of the electrons 
and $ \beta = v / c $) 
enters a vacuum interaction region with length $L_{int} $ where a plane 
monochromatic electromagnetic linearly polarized wave of laser photons with 
wavelength $ \lambda = 2\pi c / \omega $ is propagating in the 
same direction. Just as in the case of particle acceleration, 
due to the difference between the light and electron velocities, the transversal 
deflection acquired by the electrons when the phase of the waves 
acting on the electrons are varied 
less than $\pi $, will be compensated by the opposite deflection during the 
next $\pi $ variation of the phase. However, one can show that if  
$L_{int} $ is much less than $\lambda \gamma^2 $, and the difference between the 
entrance and exit phases is less than $\pi$ the 
electrons will "feel" an almost constant electric field $E $ and will 
be deflected under certain deflection angle $\alpha $ depending upon the 
entrance phase. If the electric field of the traveling wave is close to its 
maximal, amplitude value $ E \simeq E_0 $ and the magnetic field $H \simeq 0 $ 
than due to a short interaction time $t_{int} = L_{int} / 2 c $ assuming 
$ E \simeq const $ ams $ H \simeq 0$ one can show that the maximal deflection 
angle of the electrons is given by the formula        
\begin{equation}
\label{AA}
\alpha_{max} \simeq \frac{ eE \lambda \gamma}{2 mc^2 \beta} = 
\pi \gamma \xi,
\end{equation}
where $\xi = eE / mc\omega $. From (1) it follows that it is better to have 
larger wavelength ($CO_2$ laser with 
$\lambda = 10 \mu m $), relatevistic electron beams, $\gamma \gg 1$.
 or $\beta \ll 1 $. It also follows that in the case of nonrelativistic 
electron beams which usually have larger angular divergence the beam 
deflection is not realistic because $L_{int} $ is small equal to the 
wavelength. If there is a screen on a distance $L_{sc} $ after the 
interaction region then the electron beam will oscillate with 
maximal deflection $ R_{max} = L_{sc} \alpha $. 

\indent
For a photon beam  with $E \simeq 6.3.10^4 V/cm $, which is provided 
by a relatively weak laser beam with intensity $W \simeq 10^9 W/cm^2$, 
$\lambda = 10 \mu m $, ($CO_2$ laser beam with $\xi \simeq 0.0002 $) and for 
relativistic electrons with $\gamma = 100 $ ($\varepsilon $= 50 MeV) 
one obtains $\alpha = 0.06 $ radian which is much larger than the 
angular divergence of such electron beams. Therefore taking 
$L_{int} = 5 cm  \lambda \gamma^2 /2 $ and $L_{sc} = 50$ cm the 
continuous electron beam 
will oscillate on the screen with an amplitude $R = 30 $. 
The use of circularly polarized plane wave instead of linearly polarized wave 
or the addition of a second  such a perpendicularly deflecting laser system 
with appropriate phase matching will sweep the contineous electron beam 
on a circle with a lenth $L \simeq 2 \pi R = 18.85 $ cm 
and with a period $T = \lambda / c \simeq 33 $ fs. 
If the length $T_e $ of the electron beam is less than $T $ then only a part 
$ T_e / T $ of the circle will bombarded by electrons and shine. Measuring the 
length of the shining one can measure the length of the short electron pulses 
down to ten fs with accuracy of a few fs, i.e. construct femtosecond 
oscilloscopes.

\indent
It is worthwhile to make some remarks. It is well known [10] that in the 
electron rest frame the trajectories have the shapes of "eight" and circle in 
the cases of linearly and circularly polarized electromagnetic waves, 
respectively. The transversal sizes of the trajectories are small and are 
of the order of wavelength in the case of very strong fields equal to the 
critical ones $E_{crit} =m^2 c^3/ e \hbar = 1.32 10^{16} $ V/cm. The relatively 
large sizes of the figures on the screen is due to the angles (1) large 
with respect of the angular spread of the existing electron beams and 
large $L_{sc}$.The second remark concerns the velocity of the expansion of the 
shining of the screen figures which exceeds the light velocity in vacuun as 
ussually takes place in such devices.         

\indent
Here we shall not consider the sufficiently high sensitivity of the proposed 
osciloscopes as well as the influence of various factors on the amplitude and 
time measurement errors. Let us only note that due to the very short times the 
number of electrons involved into the process of deflection is very low. 
Therefore the usual screens are not suitable for detecting the arcs and circles. 
Scintillation and other types detectors with sufficient mosaity which can 
detect single electrons or secodary electrons from the process under 
investigation [11,12] can serve for detecting the figure on the screen. 
However, in this case, as it is well known 
[6,11], the maximal attainable time resolution, $\sim 10^{-14} $ s, is 
determined mainly by the spread of the initial velocities and by the 
emission time of secondary electrons coming out from a thin layer of the 
electron source. Therefore the expected time resolution is slightly less than 
10 fs, which is by one and two order order better than it is expected and 
achieved with the help of other methods.   

\indent
What concerns the other application, namely, the production of SR femtosecond 
pulses the advantage of the 
proposed method in this case is in the fact that no insertion device in the 
interaction region is required. The proposed femtosecond sweeping of the electon 
beams, of course, can find many other applications as for shortening the electron 
beams by the well known chopping methods. These applications and more correct 
consideration of the problems will be given in an other publication. One of the 
authors (K.A.I) thanks H. Avetissyan and A. Margaryan for discussions. This work 
has been supported partially by ISTC A372.

\end{document}